\documentclass[a4paper,fleqn,usenatbib]{mnras}

\usepackage[T1]{fontenc}
\usepackage{ae,aecompl}

\usepackage[normalem]{ulem}
\usepackage{amsmath}
\usepackage{graphicx} 
\usepackage{lscape}
\usepackage{indentfirst}
\usepackage{enumitem}
\usepackage{xspace}

\usepackage{amssymb}
\usepackage[dvipsnames]{xcolor}
\usepackage{url}
\usepackage[flushleft]{threeparttable}

\newcommand {\bc}{\begin {center}}
\newcommand {\ec}{\end {center}}
\newcommand {\be}{\begin {equation}}
\newcommand {\ee}{\end {equation}}
\newcommand {\beq}{\begin {eqnarray}}
\newcommand {\eeq}{\end {eqnarray}}
\newcommand {\comment}[1]{}

\title[Magnetospheric depolarisation in XRPs]
{
Magnetospheric birefringence breaks the rotating-vector model and depolarises X-ray pulsars
}
\author[A.A.~Mushtukov et al.] 
{Alexander A. Mushtukov,$^{1,2}$\thanks{E-mail: a.mushtukov@ucl.ac.uk (AAM)}
Sergey S. Tsygankov,$^{3}$
Silvia~Zane,$^{1}$
Nabil~Brice,$^{4}$
Ruth~M.~E.~Kelly,$^{5,1}$
\newauthor
Roberto~Taverna,$^{5}$
Roberto~Turolla$^{5,1}$
\\ 
$^1$ Mullard Space Science Laboratory, University College London, Holmbury St. Mary, Surrey RH5 6NT, UK\\
$^2$ Astrophysics, Department of Physics, University of Oxford, Denys Wilkinson Building, Keble Road, Oxford OX1 3RH, UK\\
$^3$ Department of Physics and Astronomy,  FI-20014 University of Turku, Finland \\
$^4$ Centre for Astrophysics Research, University of Hertfordshire, College Lane, Hatfield AL10 9AB, UK\\
$^5$ Department of Physics and Astronomy, University of Padova, via Marzolo 8, I-35131 Padova, Italy
} 

\pubyear{2026}

\begin{document}
\label{firstpage}
\pagerange{\pageref{firstpage}--\pageref{lastpage}}
\maketitle

\begin{abstract}
Accreting X-ray pulsars have long been expected to show large linear polarisation below the cyclotron resonance. 
In the standard picture, this follows from two facts: radiative opacities in the strongly magnetised neutron star environment depend strongly on polarisation mode, and the emission comes from compact polar regions. 
The emerging radiation should therefore be locally polarised and add almost coherently, with a polarisation angle tracing the projected magnetic field as in the rotating-vector model.  
Recent X-ray polarimetric observations, instead, show modest polarisation fractions and position angle swings that often depart from this simple picture. 
Here we show that propagation through the magnetosphere of an accreting neutron star can naturally produce both effects. 
Using general-relativistic ray tracing and Stokes-parameter transport, we find that photons emitted from compact polar regions acquire ray-dependent polarisation angle offsets in two ways. 
First, nearly field-aligned trajectories can cross local non-adiabatic pockets inside the QED birefringent region. 
Second, photons that intersect the plasma-loaded accretion flow accumulate additional phase shifts between plasma normal modes. 
Both effects broaden the distribution of polarisation angles across the ray bundle and reduce the hotspot-integrated linear polarisation, while also producing systematic deviations from rotating-vector-model expectations. 
Magnetospheric propagation should therefore be treated as part of the polarimetric transfer problem in accreting X-ray pulsars, and may help explain the observed low polarisation fractions and anomalous angle swings.
\end{abstract}


\section{Introduction}
\label{sec:Intro}

X-ray pulsars (XRPs; see \citealt{2022arXiv220414185M} for a review) are neutron stars (NSs) with strong magnetic fields that channel accreting matter towards the magnetic poles, producing pulsed X-ray emission. 
Their surface magnetic fields, typically $B\sim10^{11}$–$10^{13},{\rm G}$, are inferred mainly from cyclotron resonant scattering features \citep{2019A&A…622A..61S}. In such strong fields, the interaction of radiation with plasma depends strongly on photon energy, polarisation, and propagation direction \citep[see e.g.][]{2006RPPh…69.2631H,2014PhyU…57..735P,1992herm.book…..M}. 
XRPs therefore provide a natural laboratory for studying radiative processes in strongly magnetised plasmas.

Standard models predict strongly linearly polarised X-ray emission from XRPs, with polarisation degrees reaching $\sim80\%$ \citep{1985ApJ...298..147M,1985ApJ...299..138M,2021A&A...651A..12S}, particularly below the cyclotron energy,
\beq
E_{\rm cyc}\simeq 11.6
\left(\frac{B}{10^{12}\,{\rm G}}\right)\,{\rm keV}.
\eeq
This prediction became testable with modern X-ray polarimetry, most notably with the Imaging X-ray Polarimetry Explorer (IXPE; \citealt{2021AJ....162..208S}).
However, IXPE observations of several XRPs revealed much lower polarisation degrees of only $\sim5$--$20\%$ \citep{2022NatAs...6.1433D,2022ApJ...941L..14T,2024Galax..12...46P,2024NatAs...8.1047H,2025A&A...698A..22L}, in tension with standard theoretical expectations.
Part of this discrepancy may originate in the emitting layers themselves. 
For example, overheated upper atmospheric layers \citep{2018A&A...619A.114S} can reduce the emerging polarisation degree in the sub-critical regime \citep{2021MNRAS.503.5193M,2025A&A...696A.224F}.
Nevertheless, the origin of the low polarisation degrees observed in XRPs over a wide range of luminosities remains unclear.

Phase-resolved polarimetry provides additional information through the pulse-phase variation of the polarisation position angle, which we denote by $\chi$.
A commonly used description is the rotating-vector model (RVM; \citealt{1969ApL.....3..225R}), which relates the variation of $\chi$ to the orientation of the magnetic axis and is widely used to constrain the geometry of NS rotation.
Recent IXPE observations, however, have revealed energy-dependent and abrupt changes in $\chi$ in several XRPs \citep{2024NatAs...8.1047H,2025A&A...698A..22L}.
This suggests that propagation effects can modify the observed $\chi$ curve and lead to deviations from the RVM, potentially biasing geometrical constraints inferred from polarimetric data.

For photon propagation near strongly magnetised NSs, a key effect is QED vacuum birefringence \citep{1974JETP...38..903G}. 
In the usual picture of propagation through a magnetized QED vacuum, birefringence causes normal polarization modes to evolve adiabatically out to a polarization-limiting radius \citep{2000MNRAS.311..555H,2003MNRAS.342..134H}. 
The polarisation direction then freezes out and reaches the observer with a memory of the $B$-field direction at the last stage of adiabatic propagation. 
In magnetars, this effect usually aligns the polarisation vectors of photons originally emitted at different positions on the stellar surface, and therefore makes the net polarisation degree observed at infinity close to that at emission \citep{2015MNRAS.454.3254T,2015RPPh...78k6901T}.

However, the QED adiabatic region is not uniform. 
Although close to the NS the polarisation should follow the local direction of magnetic field, this is not necessarily true everywhere along the photon trajectory. 
A breakdown occurs when the photon momentum becomes nearly parallel to the field. 
In this limit the perpendicular component of the field, $B_\perp$, becomes small, the birefringent splitting is suppressed, and a local non-adiabatic pocket can appear inside the otherwise adiabatic region. 
Such quasi-tangential propagation was discussed previously by \citet{2009MNRAS.398..515W}. 
Here we show that this effect becomes especially important for XRPs because their X-ray emission originates from compact regions located close to the magnetic poles. 
We find that photons emitted from these regions are particularly sensitive to local breakdowns of adiabatic tracking, and that the resulting ray-dependent offsets of the polarisation angle can substantially modify the expected polarisation degree of the hotspot-integrated signal.

The occurrence of the non-adiabatic pockets does not require the presence of matter in the magnetosphere: they arise purely from the geometry of photon propagation through the QED-birefringent region. 
Inside the magnetospheric radius of XRPs, however, there are also regions filled with plasma supplied by the accretion disc and channelled along $B$-field lines towards NS surface. 
Within these regions, the optical properties of the medium can be dominated by plasma rather than by the QED vacuum. 
The normal modes are then elliptically polarised, rather than the linearly polarised ordinary and extraordinary modes of the matter-free propagation picture. 
This effect does not require the region to be optically thick and can operate even in the absence of scattering. 
A photon crossing the magnetospheric accretion flow can change its polarisation state by accumulating a relative phase between the plasma modes. 
The plasma-loaded accretion flow therefore provides a second region, specific to accreting sources, where the simple picture of adiabatic propagation through a magnetised vacuum can break down.

In this work, we use general-relativistic (GR) ray tracing and Stokes-parameter transport to investigate these propagation effects in accreting strongly magnetised NSs. 
We show that QED non-adiabatic pockets and plasma birefringence in the magnetospheric accretion flow can reduce the hotspot-integrated linear polarisation and produce systematic, phase-dependent departures from the RVM. 
We explore how these effects depend on photon energy, magnetic field strength, the size and geometry of the emitting region, and the viewing geometry. 
We examine the conditions under which the accretion flow modifies the polarisation signal through plasma birefringence, and assess when scattering and absorption can be neglected. 

\section{Model}
\label{sec:model}

\begin{figure*}
\centering 
\includegraphics[width=13.cm]{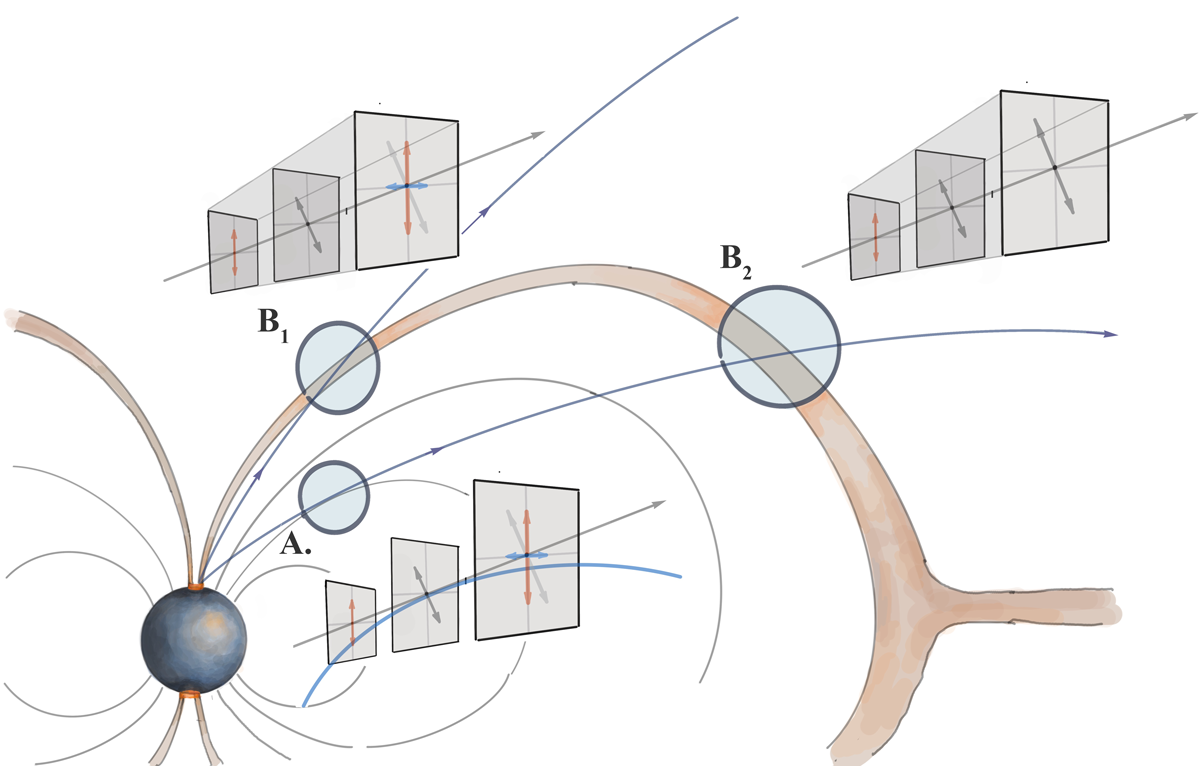}
\caption{
Schematic view of magnetospheric birefringence in an accreting X-ray pulsar.
Photons emitted from the polar regions propagate through the birefringent magnetosphere before reaching the observer.
In the standard vacuum picture, the polarisation follows the local magnetic field direction adiabatically until it freezes out.
This picture can fail in two characteristic situations, illustrated here. 
(A) First, a photon propagating nearly parallel to the magnetic field can cross a local non-adiabatic pocket inside the QED adiabatic region.
The polarisation then temporarily decouples from the instantaneous eigenmodes and acquires an offset relative to the projected magnetic field direction. 
When adiabatic tracking resumes, this offset contributes to depolarisation.
(B) Second, a photon can  cross the accretion flow. This may happen inside the QED adiabatic region (case B$_1$), in which case the flow rotates the polarisation state relative to the local vacuum eigenmodes.
After the photon leaves the flow, projection back onto the QED eigenmodes produces mode mixing and additional depolarisation.
Alternatively, (case B$_2$), if  the flow is crossed after polarisation freeze-out, the plasma layer mainly acts as a birefringent retarder: it rotates or otherwise modifies the polarisation state of that ray, but does not by itself produce single-ray depolarisation.
}
\label{pic:scheme}
\end{figure*}

We compute phase-resolved X-ray polarisation by propagating photons through a birefringent magnetosphere. 
The calculation includes both QED vacuum birefringence and plasma birefringence \citep{2000MNRAS.311..555H,2003MNRAS.342..134H} in the magnetospheric accretion flow.

We describe the results in terms of the angle $\psi$ between the line of sight and the magnetic axis. 
For a given viewing geometry
\beq
\label{eq:viewing_angle}
\cos\psi = \cos\zeta\cos\xi
+
\sin\zeta\sin\xi\cos\gamma,
\eeq
where $\zeta$ is the angle between the line of sight and the rotation axis, and $\xi$ is the angle between the magnetic and rotation axes.
The basic ingredients of the model are summarised below. 
Further technical details of the ray tracing, magnetic-field configurations, polarisation transport, accretion-flow prescription, and numerical implementation are given in the Appendices.

\subsection{Photon trajectories}
\label{sec:ray_tracing}

We account for GR light bending by propagating photons along null geodesics in the Schwarzschild spacetime. 
This provides a good approximation for the slowly rotating NSs considered here, for which rotational corrections to the spacetime are small.
At each point along a ray, the photon energy measured by a stationary observer is
\beq
E(r)=E_{\infty}
\left(1-\frac{r_{\rm s}}{r}\right)^{-1/2},
\label{eq:redshift}
\eeq
where $r_{\rm s}=2GM/c^2$ is the Schwarzschild radius and $E_{\infty}$ is the photon energy measured at infinity.

Photon trajectories are integrated numerically and embedded in three dimensions to connect the NS surface with the distant image plane.
We use a two-pass procedure: rays are first traced backwards from the image plane to the NS surface and then propagated forwards to the observer while evolving the Stokes parameters through the birefringent magnetosphere.
Further details are given in Appendix~\ref{app:ray_tracing}.

\subsection{Emission region and intrinsic polarisation}
\label{sec:emission_model}

We consider X-ray emission produced close to the magnetic poles of the NS, where the accreting matter is channelled by the magnetic field onto the stellar surface.
In the simplest case, the emitting region is represented by a circular polar cap with angular half-opening angle $\theta_{\rm max}$. 
For disc-fed accretion, its characteristic size is set by the magnetospheric radius and, for a dipolar field, can be estimated as
\beq\label{eq:theta_cap}
\theta_{\rm max}
\simeq
\left(\frac{R}{R_{\rm m}}\right)^{1/2},
\eeq
where $R$ and $R_{\rm m}$ are the NS and magnetospheric radii, respectively.
In addition to circular caps, we consider ring-like and azimuthally asymmetric footprints to explore how the source geometry affects the observed polarisation.

Our aim is to isolate the changes in polarisation produced during photon propagation through the magnetosphere rather than those arising at the emission site. 
We therefore assume that the surface radiation is initially $100\%$ linearly polarised in the X-mode,
\beq
p_{{\rm l},i}=1.
\eeq
This choice does not restrict the results to intrinsically fully polarised emission. 
Since the Stokes transport and image-plane integration are linear, a lower initial polarisation degree can be approximately accounted for by rescaling the initial $Q$ and $U$ Stokes parameters while keeping the same local polarisation-angle pattern.

We focus on compact hotspots characteristic of the sub-critical accretion regime (see e.g., \citealt{1976MNRAS.175..395B}), rather than extended radiation-dominated accretion columns. 
The intrinsic beam pattern can nevertheless affect the strength of the propagation effects: a pencil-like beam gives more weight to nearly field-aligned rays, for which the non-adiabatic effects considered here are strongest, while a fan-like beam suppresses their relative contribution.

Further details of the magnetic-field configurations and emitting-region geometries are given in Appendix~\ref{app:magnetic_field}.

\subsection{QED vacuum birefringence and polarisation transport}
\label{sec:qed_transport}

Vacuum birefringence splits the radiation into two polarisation eigenmodes with different refractive indices \citep{2000MNRAS.311..555H,2003MNRAS.342..134H}. 
When the propagation is adiabatic, the photon remains in the same eigenmode and its polarisation follows the changing direction of the local magnetic field. 
We characterise the strength of vacuum birefringence by the vector $\boldsymbol{\Omega}$, whose magnitude is
\begin{equation}
\label{eq:Omega}
|\boldsymbol{\Omega}|
=
\frac{2}{15}
\frac{\alpha_{\rm f}}{4\pi}
\frac{\omega}{c}
\left(
\frac{B_{\perp}}{B_{\rm QED}}
\right)^2
\simeq
3.9\times10^{3}
E_{\rm keV}
\left(
\frac{B_{\perp}}{B_{\rm QED}}
\right)^2
\ {\rm cm^{-1}},
\end{equation}
where $\alpha_{\rm f}$ is the fine-structure constant, $\omega$ is the photon pulsation, $B_{\rm QED}\simeq4.414\times10^{13}\,{\rm G}$ is the critical QED field, and $B_{\perp}$ is the magnetic-field component perpendicular to the photon direction.

Following \citet{2000MNRAS.311..555H,2003MNRAS.342..134H}, we treat the propagation as locally adiabatic when
\beq
\label{eq:adiabaticity}
\left|
\hat{\boldsymbol{\Omega}}
\left(
\frac{1}{|\hat{\boldsymbol{\Omega}}|}
\frac{\partial |\hat{\boldsymbol{\Omega}}|}{\partial \ell}
\right)^{-1}
\right|
\gtrsim 0.5,
\eeq
where $\ell$ is the path length along the ray and $\hat{\boldsymbol{\Omega}}$ is the unit vector along $\boldsymbol{\Omega}$.
The Stokes parameters are evolved along each ray, allowing the propagation to switch between adiabatic and non-adiabatic regimes.

The non-adiabatic regions central to this work arise when a photon propagates nearly along the $B$-field, $\theta_{kB}\ll1$. 
In this case $B_{\perp}\rightarrow0$, strongly suppressing $|\boldsymbol{\Omega}|\propto B_{\perp}^{2}$ and temporarily breaking the adiabatic tracking of the local eigenmodes. 
This temporarily breaks adiabatic tracking and produces ray-dependent polarisation-angle offsets, which can reduce the hotspot-integrated polarisation.

Further details of the numerical polarisation transport and the properties of these non-adiabatic ``pockets'' are given in Appendices~\ref{app:qed_transport} and \ref{app:pockets}.

\subsection{Birefringence in the magnetospheric accretion flow}
\label{sec:plasma_flow}

In addition to QED vacuum birefringence, photon polarisation can be modified when the radiation crosses the magnetospheric accretion flow.

We consider disc accretion onto a magnetised NS. The disc is truncated at the magnetospheric radius $R_{\rm m}$, where the plasma is captured by the magnetic field and channelled towards the magnetic poles. We represent this flow as a plasma-loaded bundle of dipolar field lines connecting the inner disc to the stellar surface. Its density is calculated from the mass accretion rate and mass continuity along the flow. Further details of the geometry and density distribution are given in Appendix~\ref{app:accretion_flow}.

Inside the accretion flow, the plasma changes the local polarisation eigenmodes and their refractive indices. These plasma modes are generally elliptically polarised and differ from the nearly linear QED eigenmodes. The two modes propagate at different phase velocities and accumulate a relative phase as the photon crosses the flow. We characterise this effect by the beat length $l_{\rm beat}$, over which the relative phase changes by $2\pi$. For a path corresponding to $N_{\rm beat}$ beat lengths,
\beq
\Delta\Phi = 2\pi N_{\rm beat}.
\label{eq:phase_lag}
\eeq
Because the density, magnetic field, and crossing geometry differ between rays, the accumulated phase shift is also ray dependent.

The effect on the observed polarisation depends on where the flow is crossed. If the crossing occurs after QED polarisation freeze-out, the plasma mainly modifies the polarisation state acquired during the preceding vacuum propagation. If the crossing occurs while QED propagation is still adiabatic, the modified state is subsequently projected back onto the QED eigenmodes. In both cases, different rays emerge with different polarisation states, which can reduce the net polarisation after integration over the emitting region.

Our accretion flow model is designed to isolate this birefringent effect rather than to provide a full radiative-transfer calculation. Photon scattering is considered separately in Appendix~\ref{app:scattering}. Unless stated otherwise, we assume that the plasma covers the entire loaded dipolar surface, maximising the fraction of rays that cross the flow. 
A smaller or azimuthally fragmented filling factor would reduce or redistribute the affected region.

\subsection{Synthetic observables}
\label{sec:observables}

The observed signal is calculated on a distant image plane. 
After propagating the rays through the magnetosphere, we express the Stokes parameters of all image-plane pixels in a common Cartesian reference frame and sum them to obtain the total Stokes vector
\beq
(I_{\rm tot},Q_{\rm tot},U_{\rm tot}).
\eeq

The linear polarisation degree $p_{\rm l}$ and the polarisation position angle $\chi$ are then
\beq
\label{eq:pl_chi}
p_{\rm l} =
\frac{\sqrt{Q_{\rm tot}^{2}+U_{\rm tot}^{2}}}{I_{\rm tot}},
\qquad
\chi = \frac{1}{2}\operatorname{atan2}(U_{\rm tot},Q_{\rm tot}),
\eeq
with the usual $180^{\circ}$ ambiguity in the linear polarisation angle; here $\operatorname{atan2}(a,b)$ returns the angle (in radians), between the positive  $x$--axis and the ray from the origin to the point $(b,\,a)$ in the Cartesian plane.

To quantify deviations from the rotating-vector model (RVM; \citealt{1969ApL.....3..225R}), we define
\beq
\label{eq:delta_chi}
\Delta\chi \equiv \chi-\chi_{\rm RVM},
\eeq
where $\chi_{\rm RVM}$ is calculated with the RVM for the same instantaneous orientation of the magnetic axis and the line of sight. 
We report the equivalent value of $\Delta\chi$ in the interval $[-90^{\circ},90^{\circ}]$.

For an axisymmetric emitting region, the observable polarisation can be parametrised by the instantaneous viewing angle $\psi$. 
For an azimuthally asymmetric footprint, it also depends on the azimuthal orientation of the emitting region relative to the observer and therefore on the full rotational geometry.

Control calculations without birefringence use the same ray tracing, emission geometry, and image-plane integration, allowing us to isolate propagation-induced changes from purely geometrical effects.

\section{Results}
\label{sec:results}

We first consider a low-density magnetosphere and focus on the effect of QED vacuum birefringence alone. This provides a baseline for understanding where the standard picture of adiabatic propagation can fail, before we turn to the inclusion of regions filled with the accretion flow.

\subsection{QED vacuum birefringence}
\label{sec:results_qed}

To separate propagation effects from simple geometric averaging, which occurs when the emission region is extended, we consider a circular polar emitting region of latitudinal semi-amplitude $\theta_{\max}$ and compare calculations performed, for the same model parameters, with and without accounting for QED vacuum birefringence.

\begin{figure*}
\centering
\includegraphics[width=16cm]{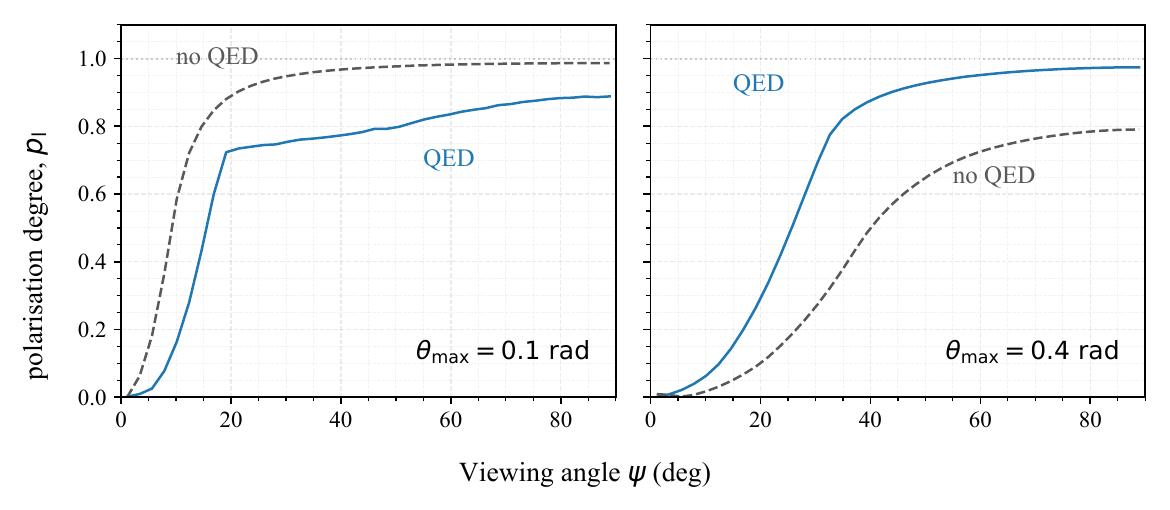}
\caption{
Comparison of the phase-resolved linear polarisation degree $p_{\rm l}$ computed with (solid) and without (dashed) accounting for QED vacuum birefringence effects in the low-density magnetosphere.
If emission originates in a small region (left panel), local non-adiabatic pockets imprint ray-dependent polarisation-angle offsets, broadening the distribution of polarisation angles across the image plane. Thus, when the Stokes parameters are summed to reconstruct the signal integrated over the emitting region, QED effects produce a reduction in $p_{\rm l}$.
On the contrary, if the emitting region is extended (right panel), the depolarisation induced by surface averaging dominates in the no-QED case, while birefringence partially re-aligns the polarisation vectors.
Here a dipole field with polar strength $B_{\rm p}=10^{12}\,\mathrm{G}$ is assumed; the photon energy is $E=1\,\mathrm{keV}$ and seed photons are $100\%$ polarised in X-mode.
The emitting-region sizes are $\theta_{\max}=0.1\,\mathrm{rad}$ and $\theta_{\max}=0.4\,\mathrm{rad}$ in the left and right panels, respectively.
}
\label{pic:QED_compare}
\end{figure*}

Figure~\ref{pic:QED_compare} shows two representative cases for a dipolar magnetosphere and a photon energy $E=1\,\mathrm{keV}$.
In the case of emission from a compact hotspot, the calculation without QED predicts a highly polarised signal, because the Stokes vectors from different rays add nearly coherently at infinity due to the small size of the emitting region.
However, when QED birefringence is included, the expected linear polarisation degree is reduced.
This is caused by the presence of local non-adiabatic pockets, which introduce ray-dependent polarisation-angle offsets across the image plane.
These offsets broaden the distribution of polarisation angles of the emitted photons, so that $p_{\rm l}$, the net polarisation degree integrated over the hotspot area, is reduced.

For a larger emitting region, the behaviour is closer to the familiar magnetar-like case.
Without QED, the signal is partially depolarised as the result of surface averaging.
Vacuum birefringence can then partially realign the polarisation vectors before freeze-out, increasing the expected net $p_{\rm l}$ relative to the no-QED calculation.
QED vacuum birefringence can therefore either enhance or decrease the expected polarisation degree, and this depends on the size of the emitting region.

For a dipolar field, the transition occurs near a characteristic half-opening angle $\theta_{\rm crit}$, which depends on photon energy and magnetic-field strength.
In the parameter range relevant to accreting XRPs, $\theta_{\rm crit}$ is of the order of a few degrees.
An analytic estimate of $\theta_{\rm crit}(E,B)$, together with a broader exploration of the dependence on photon energy, magnetic-field strength, emitting-region geometry, field configuration and NS compactness, is given in Appendix~\ref{app:parameter_dependence}.


The efficiency of the QED pocket effect in decreasing the polarisation degree becomes weaker at higher photon energies and for higher magnetic-field strengths, since in these cases the QED-induced realignment of the photon polarisation to the local eigenmodes is stronger and the pocket-induced offsets are smaller.
The geometry of the emitting region also matters: photons originating from small filled caps are more strongly depolarised than those emitted from larger or ring-like regions (see again  Appendix~\ref{app:parameter_dependence} for more details).

\subsection{Polarisation-angle deviations from the rotating-vector model}
\label{sec:results_pa}


For an azimuthally symmetric emitting region, the Stokes-vector summation remains symmetric around the magnetic axis.
In this case, the net polarisation angle remains very close to the RVM prediction, even if the linear polarisation degree is reduced by ray-dependent propagation effects.
A non-zero residual $\Delta\chi$ requires this azimuthal symmetry to be broken.

\begin{figure}
\centering
\includegraphics[width=\columnwidth]{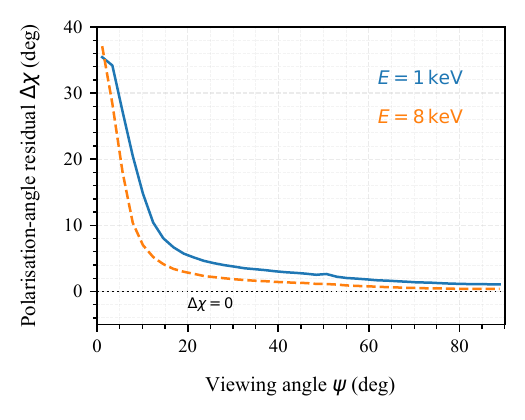}
\caption{
Deviation of the observed polarisation position angle from the RVM, $\Delta\chi=\chi-\chi_{\rm RVM}$, for an azimuthally asymmetric emitting region.
The dotted line marks the RVM prediction for which $\Delta\chi=0$.
The emitting region is a half-cap with $\theta_{\min}=0$ and $\theta_{\max}=0.1\,\mathrm{rad}$, with a fixed azimuthal orientation in the magnetic frame.
For an azimuthally symmetric emitting region, the corresponding residual is close to zero because the ray-dependent polarisation-angle offsets largely cancel in the Stokes sum.
In the asymmetric case shown here, this cancellation is incomplete, producing a coherent shift of the net polarisation angle.
The residual is plotted as a function of the instantaneous viewing angle $\psi$; however, for an asymmetric emitting region it can in general also depend on the azimuthal orientation of the spot relative to the observer.
Here we assume a dipole field with $B_{\rm p}=3\times10^{12}\,\mathrm{G}$, $100\%$ X-mode seed emission, and photon energies $E=1$ and $8\,\mathrm{keV}$.
}
\label{pic:pol_ang}
\end{figure}

Figure~\ref{pic:pol_ang} shows such a case, in which the emitting region is an azimuthally asymmetric half-cap.
The curves are plotted as a function of the instantaneous viewing angle $\psi$ for one fixed azimuthal orientation of the emitting region in the magnetic frame.
They should therefore be interpreted as a diagnostic example of the magnitude of the propagation-induced polarisation-angle bias, rather than as a universal additive correction to an arbitrary RVM curve.
In general, for an asymmetric emitting region, $\Delta\chi$ can depend not only on $\psi$, but also on the azimuthal orientation of the emitting region with respect to the observer, or equivalently on the full rotational geometry.

We find that the deviations are largest, up to tens of degrees, when the line of sight passes close to the magnetic axis, while they decrease at larger viewing angles and for higher photon energies.
This behaviour occurs because, for an asymmetric emitting region, the Stokes-vector cancellation between photons is unbalanced, producing a systematic shift of the net polarisation angle away from the RVM curve.
For an axisymmetric emitting region, in contrast, the pocket-induced spread of polarisation angles mainly reduces $p_{\rm l}$, while the total polarisation angle remains close to the RVM prediction.

\subsection{Plasma birefringence in the accretion flow}
\label{sec:results_plasma}

We now include the plasma-loaded accretion flow described in Section~\ref{sec:plasma_flow}. 
Photons intersecting the flow acquire ray-dependent phase shifts between the plasma normal modes, which can modify both their polarisation angle and degree.
We compare the pure-QED reference calculation with models of increasing mass accretion rate at fixed surface magnetic field.

\begin{figure*}
\centering
\includegraphics[width=0.98\textwidth]{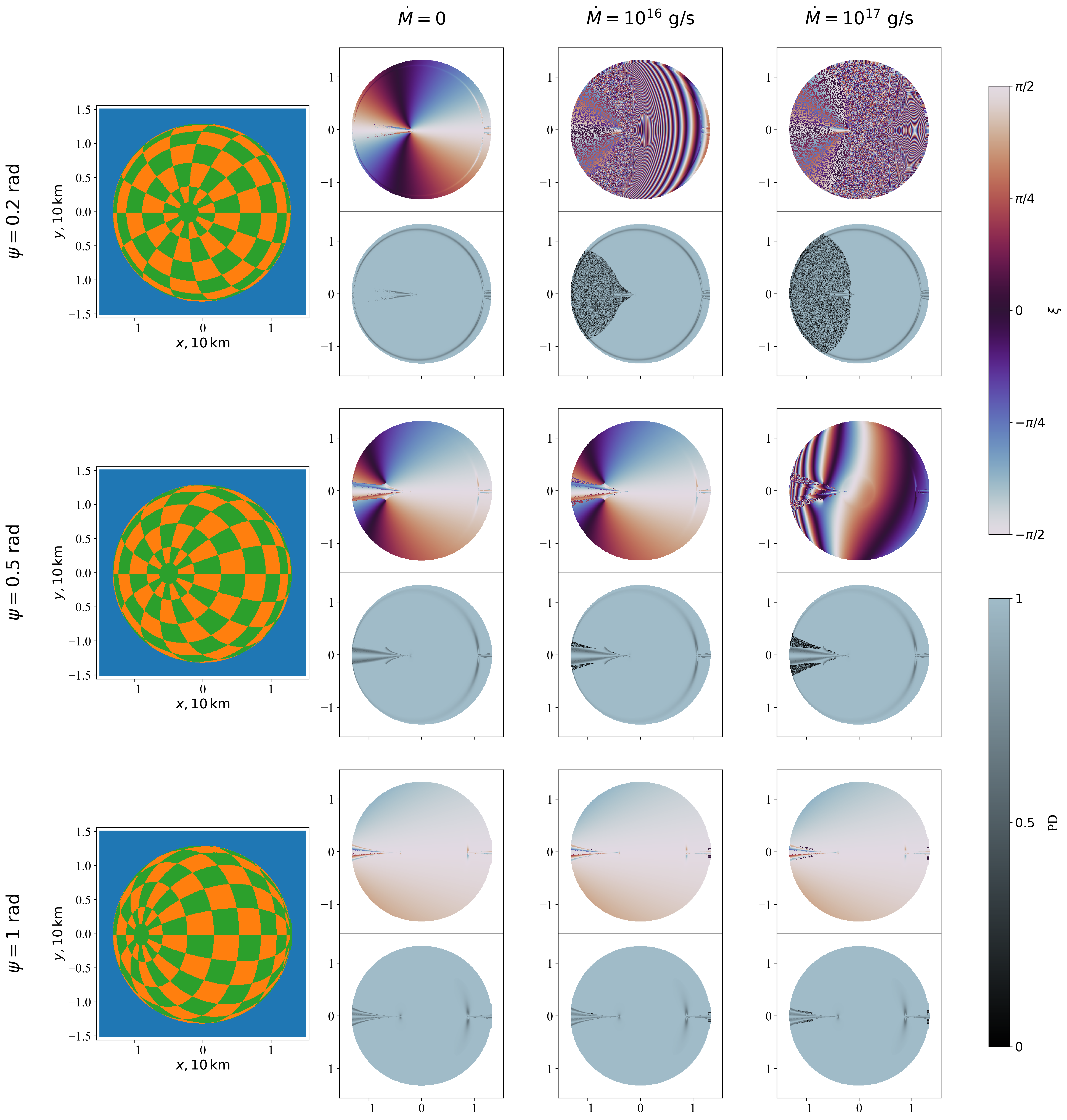}
\caption{
Image-plane polarisation diagnostics including propagation through the magnetospheric accretion flow.
The maps are computed for a dipolar magnetic field with polar strength $B_{\rm p}=10^{12}\,{\rm G}$, a NS mass $M=1.4\,M_\odot$ and radius $R=10\,{\rm km}$.
Seed photons are assumed to be $100\%$ linearly polarised in the X-mode.
Rows correspond to different viewing angles $\psi$, and columns correspond to mass accretion rates $\dot M=0$, $10^{16}$, and $10^{17}\,{\rm g\,s^{-1}}$.
For each case, the upper map shows the angle $\xi$ between the projected magnetic axis and the observed polarisation vector, while the lower map shows the propagation-induced reduction of the linear polarisation degree for photons emitted from each point of the NS surface.
The case $\dot M=0$ gives the vacuum/QED reference calculation, whereas the cases with $\dot M>0$ include the additional plasma-birefringent effect of the accretion flow.
Photons crossing the stream accumulate a phase lag between plasma normal modes, which changes the emergent Stokes vector in a ray-dependent way.
The affected image-plane region grows with increasing $\dot M$ and is especially prominent for lines of sight close to the magnetic axis.
}
\label{pic:maps_plasma_eff}
\end{figure*}

Figure~\ref{pic:maps_plasma_eff} shows the plasma-birefringent effect on the image plane.
Since both the mass density and magnetic-field strength vary across the magnetospheric stream, photons emitted from different locations encounter different local plasma conditions when crossing the flow.
The resulting plasma-induced phase shift is therefore ray-dependent, rather than a uniform rotation of the Stokes vector applied to the entire ray bundle.
The subset of rays intersecting the accretion-loaded magnetosphere acquires a sizeable plasma-mode phase lag, which causes a strongly non-uniform pattern of polarisation-angle offsets and local depolarisation across the image plane.
The extension of the affected region grows with $\dot M$ and is largest for small $\psi$, where the line of sight samples the densest part of the stream.

\begin{figure*}
\centering
\includegraphics[width=0.80\textwidth]{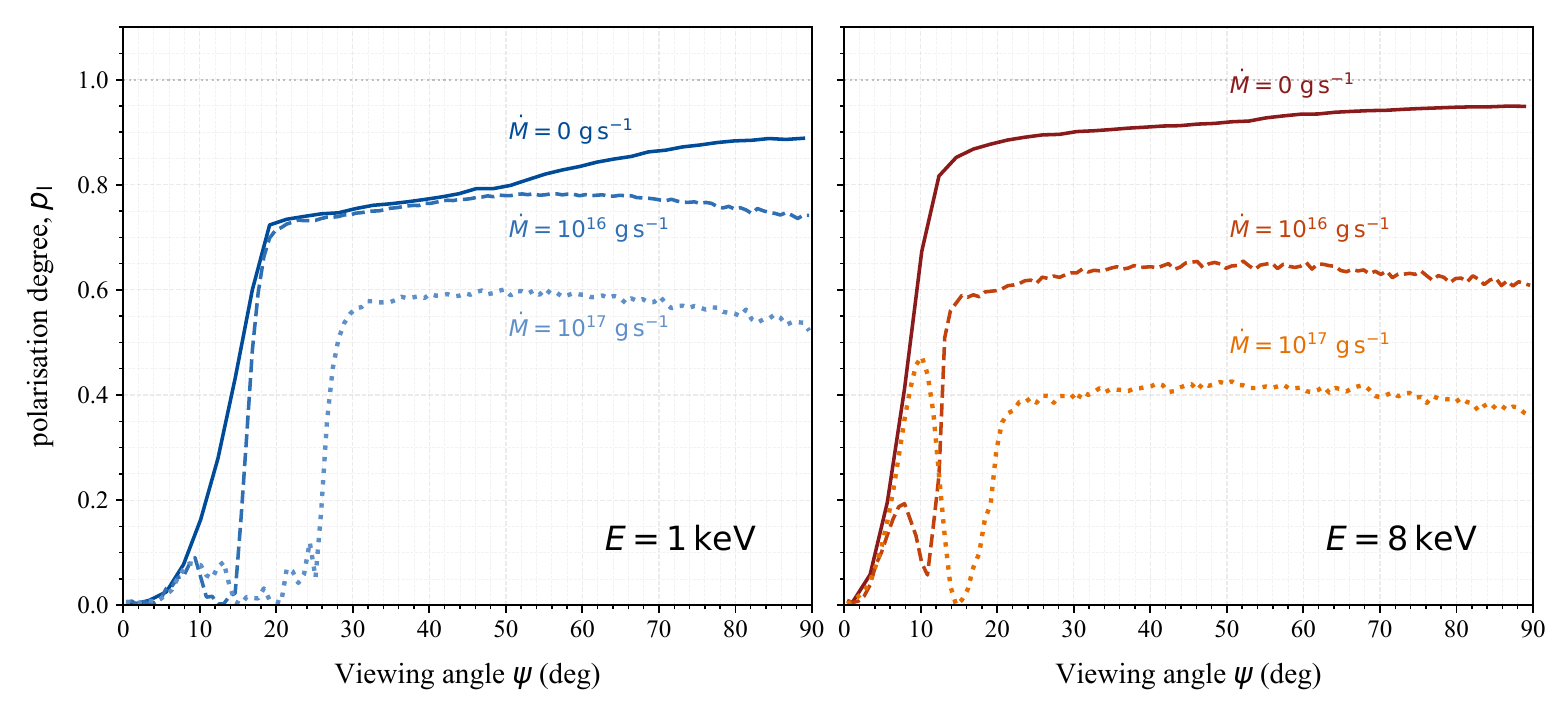}
\caption{
Effect of the magnetospheric accretion flow on the linear polarisation degree.
The curves show the hotspot-integrated $p_{\rm l}$ as a function of viewing angle $\psi$, for a polar magnetic field $B_{\rm p}=10^{12}\,{\rm G}$ and hotspot size $\theta_{\max}=0.1\,\mathrm{rad}$.
Left: $E=1\,{\rm keV}$. Right: $E=8\,{\rm keV}$.
In each panel, solid, dashed and dotted curves correspond to $\dot M=0$, $10^{16}$, and $10^{17}\,{\rm g\,s^{-1}}$, respectively.
The plasma-loaded accretion flow is assumed to be axisymmetric about the magnetic axis, so that its effect does not depend on the magnetic azimuth, as representative of the case of a purely aligned rotator.
}
\label{pic:sc_pl_Mdot}
\end{figure*}

We then consider a circular emitting polar region with $\theta_{\max}=0.1\,\mathrm{rad}$.
The corresponding hotspot-integrated observed polarisation degree is shown in Fig.~\ref{pic:sc_pl_Mdot}.
Increasing $\dot M$ reduces the phase-resolved $p_{\rm l}$, with the strongest suppression occurring at small viewing angles, where the ray bundle samples the densest part of the plasma-loaded magnetosphere.

We stress that this reduction is not a dilution effect from unpolarised scattered radiation.
Scattering is not included in the polarisation calculation presented here.
The presence of the accretion flow therefore provides a second, accretion-rate-dependent depolarisation channel in addition to the QED non-adiabatic-pocket mechanism.
The validity of neglecting scattering, in particular for lines of sight close to the magnetic axis where the propagation effects discussed above are strongest, is tested independently in Appendix~\ref{app:scattering}.

\section{Discussion and conclusions}
\label{sec:discussion}

Magnetospheric propagation of X-ray photons can modify both the linear polarisation degree (PD) and the polarisation angle (PA) of accreting XRPs.
QED vacuum birefringence does not necessarily increase the observed PD. 
For compact emitting regions, local non-adiabatic pockets produce different polarisation-angle offsets along different photon trajectories, reducing the net polarisation after integration over the hotspot. 
For typical XRP parameters, this effect operates for emitting regions with angular scales of a few degrees and is strongest at lower photon energies and for trajectories close to the magnetic axis. 
For larger emitting regions, the conventional QED alignment effect dominates.

The effect on the PA depends strongly on the symmetry of the emitting region. 
For an axisymmetric hotspot, propagation-induced offsets largely cancel and the resulting PA remains close to the RVM prediction. 
For an azimuthally asymmetric footprint, this cancellation is incomplete and deviations from the RVM can reach tens of degrees. 
Magnetospheric propagation can therefore bias geometrical constraints inferred directly from PA curves.

Plasma birefringence in the magnetospheric accretion flow provides an additional source of depolarisation. 
Its importance increases with the mass accretion rate because different rays acquire different plasma-induced phase shifts. 
This suggests an observational test: if plasma birefringence contributes significantly to the observed depolarisation, the PD should generally decrease with increasing luminosity. 
Existing IXPE observations cover only modest luminosity variations \citep{2023A&A...675A..48T}; observations of bright transient XRPs over a broader luminosity range, possible with future missions such as eXTP \citep{2025SCPMA..6819505G}, can test this prediction.

Our treatment isolates birefringent propagation and does not include scattering in the main polarisation calculation. 
The Monte Carlo test in Appendix~\ref{app:scattering} shows that this approximation is adequate at $\dot M\lesssim10^{16}\ {\rm g\,s^{-1}}$, where unscattered photons dominate over most viewing directions. 
Scattering becomes important as $\dot M$ approaches $\sim10^{17}\ {\rm g\,s^{-1}}$, close to the regime where the hotspot approximation itself breaks down and an accretion column is expected to form.

The main conclusion is therefore that the magnetosphere is not a passive propagation region. 
Both QED and plasma birefringence can alter the observed polarisation degree, while propagation through an asymmetric system can also distort the position-angle curve away from the RVM. 
These effects should be accounted for when using X-ray polarimetry to constrain the emission and magnetic geometry of X-ray pulsars.

\section*{Acknowledgements}

AAM acknowledges support from the UKRI1169 Astronomy observation and theory small award.
This research was supported by the International Space Science Institute (ISSI) in Bern, through the International Team project 25-657 `Polarimetric Insights into Extreme Magnetism' and the  Research Council of Finland Centre of Excellence in Neutron-Star Physics (grant 374064).

\section*{Data availability}

The calculations presented in this paper were performed using a private code developed and owned by the corresponding author. All the data appearing in the figures are available upon request. 



\appendix

\section{Photon trajectories and magnetic-field configuration}
\label{app:geometry}

\subsection{Photon trajectories in Schwarzschild spacetime}
\label{app:ray_tracing}

We model the exterior spacetime of the NS with the Schwarzschild metric
and compute photon trajectories as null geodesics.
Each trajectory lies in a plane and can be parametrised using polar
coordinates, with $r\geq0$ and $\phi\in[0;2\pi]$.
By introducing the inverse dimensionless radius
\beq
u \equiv \frac{r_{\rm s}}{2r},
\eeq
where $r_{\rm s}=2GM/c^{2}$ is the Schwarzschild radius, the standard
orbit equation reads (see Section~23 in \citealt{1973grav.book.....M}),
\beq
\frac{{\rm d}^{2}u}{{\rm d}\phi^{2}} = 3u^{2} - u.
\label{eq:schwarzschild_orbit_app}
\eeq

We integrate the previous equation step-by-step, updating the direction
of the photon along the path; to perform the calculations in 3D, we
follow the procedure described in \citet{2025MNRAS.538.2396M}.
This approach fully accounts for light bending and provides the mapping
between the NS surface and the observer's image plane used in the
numerical pipeline (Appendix~\ref{app:image}).

The local energy (defined in equation~\ref{eq:redshift}) is consistently
used when evaluating all quantities along each ray (e.g. the
birefringence strength; equation~\ref{eq:Omega}).

For rays that intersect the stellar surface, we construct the emission
direction in the local stationary orthonormal frame at the surface and
compute the angle between the ray and the local surface normal.
This information determines the adopted beam pattern and specifies the
intrinsic surface polarisation state used as the initial condition for
forward propagation (Appendix~\ref{app:image}).

\subsection{Magnetic-field configuration}
\label{app:magnetic_field}

We consider axisymmetric magnetic fields aligned with the stellar
rotation axis (the $z$-axis).
We use spherical coordinates $(r,\theta,\phi)$, where $\theta$ is the magnetic colatitude.
Two configurations are explored: (i) a dipolar magnetic field and
(ii) a pure axisymmetric quadrupolar magnetic field.

In the Newtonian limit, the vacuum dipole field has the standard spherical components
\beq\label{eq:Bdip_sph}
B^{\rm (d)}_{r} =
\frac{2\mu_{\rm d}\cos\theta}{r^{3}},
\qquad
B^{\rm (d)}_{\theta} =
\frac{\mu_{\rm d}\sin\theta}{r^{3}},
\qquad
B^{\rm (d)}_{\phi} = 0,
\eeq
where $\mu_{\rm d}$ is the magnetic dipole moment,
\beq
\mu_{\rm d}\equiv \frac{B_{0}R^{3}}{2}\,,
\eeq
and $B_{0}\equiv B_{\rm pol}$, with $B_{\rm pol}$ the polar surface field strength.

For the quadrupole field we use the standard axisymmetric vacuum solution in spherical components,
\beq
B^{\rm (q)}_{r} = \mu_{\rm q}\,
\frac{3\cos^{2}\theta-1}{r^{4}},
\qquad
B^{\rm (q)}_{\theta} = \mu_{\rm q}\,
\frac{\sin 2\theta}{r^{4}},
\qquad
B^{\rm (q)}_{\phi} = 0,
\label{eq:Bquad_sph}
\eeq
where $\mu_{\rm q}$ sets the quadrupole strength
(an overall normalisation).

GR corrections to the magnetic-field structure are
included using exact static solutions of Maxwell's equations in the
Schwarzschild spacetime
\citep{1992MNRAS.255...61M,2017MNRAS.472.3304P}.
We evaluate the magnetic field in the local orthonormal frame of a
static observer (FIDO) in Schwarzschild coordinates, i.e. using the
physical components $B^{\hat r}$ and $B^{\hat\theta}$ appropriate for
the curved spatial metric (see \citealt{2017MNRAS.472.3304P} for
conventions in the $3{+}1$ formalism).

\section{Vacuum birefringence, numerical implementation and
non-adiabatic pockets}
\label{app:qed_details}

\subsection{Vacuum birefringence and polarisation transport}
\label{app:qed_transport}

The propagation of polarised radiation through the magnetosphere is
affected by QED vacuum birefringence, which introduces two normal
polarisation modes with different refractive indices
\citep{2000MNRAS.311..555H,2003MNRAS.342..134H}.
To describe this effect quantitatively, we characterise the strength
of birefringence by the birefringence vector $\boldsymbol{\Omega}$
(equation~\ref{eq:Omega}), which depends on the local photon energy
and on the perpendicular magnetic-field component $B_{\perp}$.

We evolve the Stokes parameters along each ray in a parallel-transported
orthonormal basis tied to the photon momentum.
At each step, we relate the Stokes vector to the instantaneous
eigenmode basis defined by the projected field direction
$\mathbf{B}_{\perp}$ in the plane orthogonal to the photon momentum.
A fixed Cartesian reference frame on the image plane is used for the
final summation (Appendix~\ref{app:image}).

The coupling between polarisation modes depends on how rapidly the
birefringence vector varies along the ray.
Following
\citet{2000MNRAS.311..555H,2003MNRAS.342..134H},
we treat the evolution as adiabatic when the criterion given by
equation~(\ref{eq:adiabaticity}) is satisfied and as non-adiabatic
otherwise.
When the adiabaticity condition holds, the photon is assumed to remain
in the same instantaneous normal mode (no mode coupling), i.e. the
mode amplitudes are conserved in the local eigenmode basis tied to
$\mathbf{B}_{\perp}$ while the eigenmode directions vary along the ray;
when the condition is violated, the polarisation is taken to be frozen
with respect to the propagation/transport frame (i.e. it does not
follow the instantaneous normal-mode basis tied to $B_{\perp}$), and
the Stokes parameters are transported without further mode evolution.

Since the projected magnetic-field direction can still rotate along
the ray, the photon polarisation emerging from such a segment is
generally rotated with respect to the local eigenmodes when adiabatic
tracking resumes.

Local non-adiabatic ``pockets'' arise most naturally for trajectories
with $\theta_{\rm kB}\ll1$, where $B_{\perp}\to0$ and the
birefringence strength is suppressed,
$|\boldsymbol{\Omega}|\propto B_{\perp}^{2}$.
In this limit the local eigenmode basis becomes ill-behaved: the
direction of $\mathbf{B}_{\perp}$ can change rapidly even when the
photon direction changes little along the ray.

Importantly, the depolarisation is not produced by the pocket interior
as a separate dissipative process; rather, the pocket allows the
polarisation to decouple from the instantaneous eigenmodes while the
projected field direction rotates, and the resulting mismatch is
imprinted when the ray exits the pocket and adiabatic evolution resumes.
When the Stokes parameters are subsequently summed over many rays and
surface elements, the accumulated dispersion of polarisation position
angles leads to partial cancellation and hence an additional
depolarisation channel.

A formal pocket definition and diagnostics are given in
Appendix~\ref{app:pockets}.

\subsection{Numerical pipeline and image-plane integration}
\label{app:image}

Our numerical pipeline follows the standard image-plane ray-tracing
strategy, adapted to emission from a NS surface and to the evolution of
photon polarisation in a birefringent magnetosphere.
We introduce a distant image plane perpendicular to the observer's
line of sight and discretise it into pixels, the centres of which are
at a general position $(X,Y)$.
Each pixel corresponds to a small solid-angle element, so summation
over the image plane is equivalent to integration over the observed
solid angle.
The computation proceeds in two passes.

\subsubsection{Backward pass: surface mapping}

In the first pass we trace photons backwards from the image plane
towards the NS by integrating the null geodesics in the Schwarzschild
spacetime (Appendix~\ref{app:ray_tracing}).
If a ray intersects the stellar surface, we record:
(i) the surface coordinates of the intersection point and
(ii) the emission direction in the local stationary orthonormal frame
at the surface.

In particular, we determine the angle between the ray and the local
surface normal required to evaluate the assumed surface emission
angular pattern and to set the initial polarisation state.
This backward mapping provides a visibility map of the surface
(see left panels in Fig.~\ref{pic:maps}) and establishes a one-to-one
correspondence between surface elements and image-plane pixels.

\subsubsection{Forward pass: polarisation evolution}

In the second pass we propagate rays forward from the surface to the
image plane.
We start by prescribing the specific intensity and the intrinsic
polarisation (e.g. an O-/X-mode mixture expressed as an initial Stokes
vector) over the emission region.

We then evolve the Stokes parameters along the ray in the presence of
QED vacuum birefringence, evaluating the local
$\boldsymbol{\Omega}$ (equation~\ref{eq:Omega}) and applying the
step-by-step switching prescription between adiabatic tracking and
freeze-out according to the adiabaticity criterion
(equation~\ref{eq:adiabaticity}).

After the forward pass we obtain, for each image-plane pixel, the
Stokes parameters $(I,Q,U)$ expressed in a common image-plane reference
frame.
Exploiting the additivity of the Stokes parameters, we sum the
contributions of all pixels to compute the total observed Stokes vector
$(I_{\rm tot},Q_{\rm tot},U_{\rm tot})$ and derive the net degree of
linear polarisation and the polarisation position angle via
equation~\ref{eq:pl_chi}.

In addition, when constructing polarisation maps we compute the angle
$\xi$ between the projected magnetic-axis direction and the local
polarisation vector on the image plane using the same projection
geometry for all rays.
This provides a diagnostic of how the surface-to-observer mapping,
together with QED-driven evolution and intermittent non-adiabatic
pockets, broadens the distribution of polarisation angles across the
observed emission region.

\subsection{Diagnostics of non-adiabatic pockets}
\label{app:pockets}

In our step-by-step integration, local departures from adiabatic
transport are identified through the adiabaticity criterion
(equation~\ref{eq:adiabaticity}).
We define a \emph{non-adiabatic pocket} along a given ray as a
contiguous path segment where the criterion is violated, embedded
within regions where the evolution is adiabatic.

The resulting adiabatic structure is trajectory-dependent:
adiabaticity is a property of a ray segment, not of a spatial point
alone.
A given point in the magnetosphere cannot be labelled as adiabatic or
non-adiabatic without specifying the photon direction, because both
the local birefringence strength and the rotation rate of the
eigenmode basis depend on the angle between the ray and the magnetic
field.

\begin{figure*}
\centering
\includegraphics[width=0.9\textwidth]{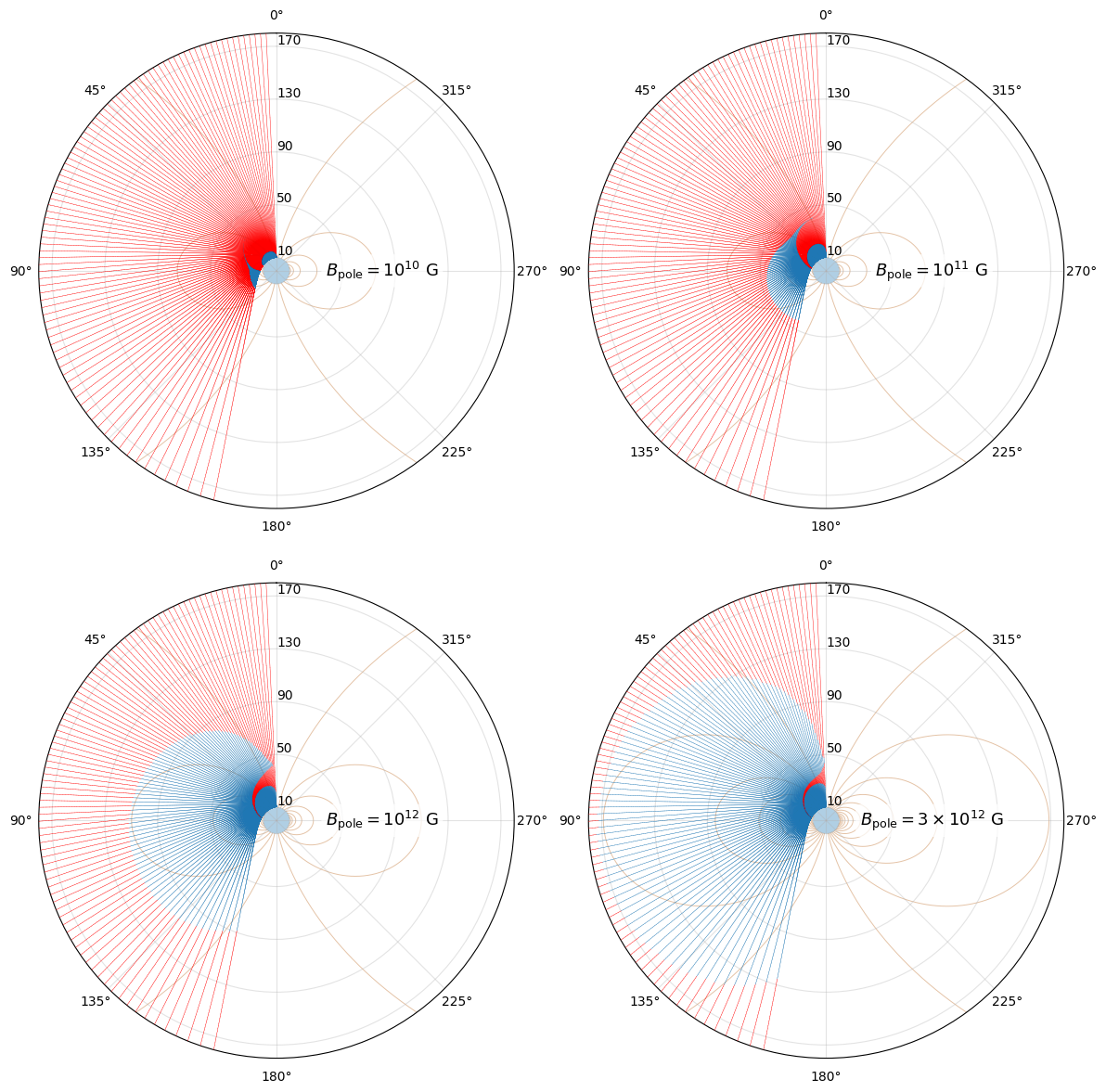}
\caption{
Trajectory-dependent geometry of the QED adiabatic and non-adiabatic
propagation domains.
The panels show trajectories of photons emitted from the magnetic pole of a NS and moving in a Schwarzschild spacetime for different polar magnetic-field strengths.
Blue segments mark regions where the local adiabaticity criterion is
satisfied and the polarisation follows the instantaneous QED eigenmodes
tied to the projected magnetic field.
Red segments mark regions where the criterion is violated and the
polarisation is effectively frozen in the transport basis.
The figure shows how the adiabatic domain is not a fixed spatial volume
of the magnetosphere: its location and extent depend on the photon
trajectory, in particular on the angle between the photon momentum and
the local magnetic field.
Non-adiabatic pockets arise naturally along ray families that become
nearly aligned with the magnetic field, where $B_\perp\to0$ and
birefringent splitting is suppressed.
}
\label{fig:adiabatic_zones}
\end{figure*}

Figure~\ref{fig:adiabatic_zones} illustrates this trajectory-dependent
geometry for photons emitted from the magnetic pole in a Schwarzschild
spacetime for different magnetic-field strengths.

\begin{figure*}
\centering
\includegraphics[width=16.cm]{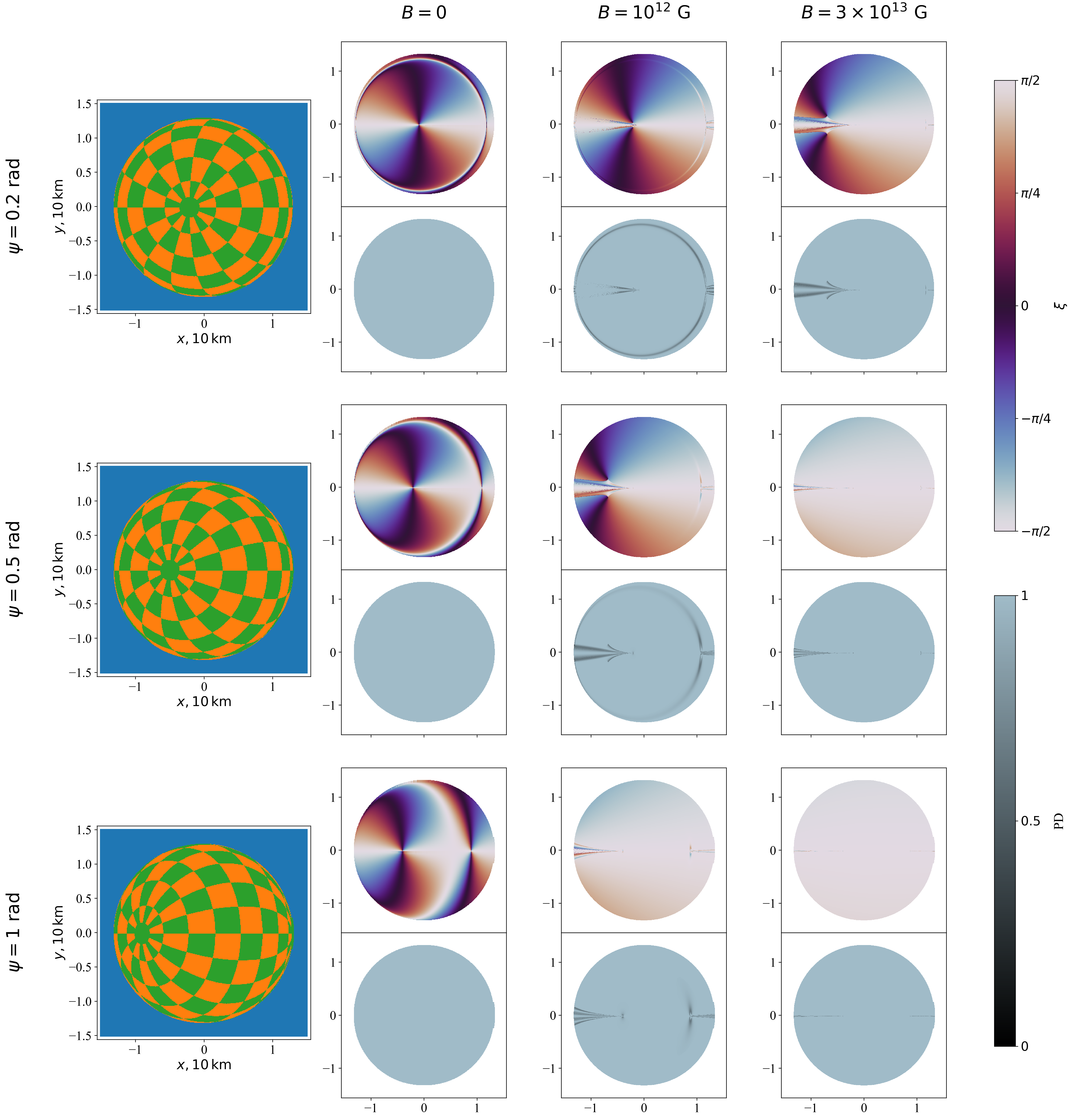}
\caption{
Image-plane polarisation diagnostics for emission from different
surface elements.
Top: angle $\xi$ between the projected magnetic axis and the observed
polarisation vector; bottom: local contribution to the reduction of
the net linear polarisation.
Increasing $B$ shifts the freeze-out outward and enhances coherence
across the image plane, except along ray families that experience local
non-adiabatic pockets (see Appendix~\ref{app:pockets}).
Columns: $B\simeq0$, $10^{12}\,\mathrm{G}$,
$3\times10^{13}\,\mathrm{G}$ (dipole; X-mode seed emission).
$M=1.4\,M_\odot$, $R=10\,\mathrm{km}$.
}
\label{pic:maps}
\end{figure*}

Pockets arise most naturally for trajectories with
$\theta_{\rm kB}\ll1$, where $B_\perp\to0$ and
$|\boldsymbol{\Omega}|\propto B_\perp^2$ is suppressed.
In this limit, the projected field direction
$\mathbf{B}_\perp$ can rotate rapidly near the pocket boundaries,
making the local eigenmode basis ill-conditioned and enhancing mode
coupling.

The physically important locations are therefore the pocket boundaries
rather than the pocket interior itself: when a ray enters or leaves a
pocket, the polarisation switches between following and not following
the instantaneous eigenmode basis.

To connect pocket locations to the image-plane diagnostics used in the
figures, we record for each ray the radial intervals where the
adiabaticity criterion is violated and mark the corresponding pixel
families on the image plane (e.g. Fig.~\ref{pic:maps}).
These trajectory-dependent switches imprint ray-dependent offsets in
the polarisation angle that persist after adiabatic tracking resumes.
After summation over the image plane, the accumulated spread of these
offsets broadens the polarisation-angle distribution and reduces the
net linear polarisation degree.

\subsection{Rotating-vector model and polarisation-angle deviations}
\label{app:rvm}

To quantify departures from the rotating-vector model
(RVM; \citealt{1969ApL.....3..225R}), we compare the net polarisation
position angle $\chi$ obtained from the summed Stokes parameters with
the angle $\chi_{\rm RVM}$ predicted for the same viewing geometry.

We adopt the standard RVM convention in which the observed polarisation
angle is set by the orientation of the projected magnetic axis on the
plane of the sky.
For the geometry used in this work, $\chi_{\rm RVM}$ is computed from
the instantaneous directions of the rotation axis, magnetic axis, and
line of sight, using the same image-plane reference axis as for $\chi$
(equation~\ref{eq:pl_chi}).

The polarisation-angle residual is defined as
$\Delta\chi\equiv\chi-\chi_{\rm RVM}$ and is evaluated modulo $\pi$;
we use the same branch choice for $\chi$ and $\chi_{\rm RVM}$ to avoid
artificial $90^\circ$ flips when analysing phase-resolved curves.

\section{Magnetospheric accretion stream and plasma birefringence}
\label{app:accretion_flow}

In accreting XRPs the magnetosphere contains plasma supplied by the
disc and channelled by the magnetic field towards the stellar surface.
Even in the absence of scattering, this magnetospheric flow can modify
the polarisation state of X-ray photons because its dielectric response
is plasma-dominated.

Here we do not solve the full radiative-transfer problem inside the
flow; instead, we estimate the phase retardation accumulated by a
photon crossing the flow and correspondingly update the Stokes vector.

\subsection{Geometry and density of the accretion flow}
\label{app:flow_geometry}

We assume disc accretion onto a magnetised NS.
The disc is truncated at the magnetospheric radius, where the stellar
magnetic field disrupts the disc flow and channels the plasma along the
field lines.

We estimate this radius as
\begin{equation}
R_{\rm m}
\simeq
3.5\times 10^8\,
\Lambda\,
B_{12}^{4/7}
\dot M_{17}^{-2/7}
\left(\frac{M}{M_\odot}\right)^{-3/7}
R_6^{12/7}\,
{\rm cm},
\label{eq:Rmag}
\end{equation}
where $B_{12}=B_{\rm p}/10^{12}\,{\rm G}$,
$B_{\rm p}$ is the polar surface magnetic field strength,
$\dot M_{17}=\dot M/(10^{17}\,{\rm g\,s^{-1}})$,
$R_6=R/(10^6\,{\rm cm})$,
$M$ and $R$ are the NS mass and radius, and
$\Lambda\sim0.5$ parametrises the coupling between the disc and the
magnetosphere.

At the inner disc edge, we assume that plasma penetrates into the
magnetosphere over a radial width comparable to the local geometrical
thickness of the disc.
At relatively low luminosity and for a gas-pressure-dominated disc, the
thickness can be estimated as
\beq\label{eq:Hd}
H_{\rm d} =
2.8\times 10^6\,\Lambda^{9/8} B_{12}^{9/14}
\dot M_{17}^{-6/35}
\left(\frac{M}{M_\odot}\right)^{-15/28}
\ {\rm cm},
\eeq
where $H_{\rm d}$ is evaluated at $R_{\rm m}$, assuming that the disc
is truncated in the C-zone \citep{2007ARep...51..549S}.
This estimate sets the transverse size of the bundle of field lines
loaded by the accreting plasma.

Inside $R_{\rm m}$ we assume that the flow is controlled by the
magnetic field and moves along the field lines.
We use magnetic spherical coordinates
$(r,\theta_{\rm m},\phi_{\rm m})$, with the polar axis aligned with the
magnetic dipole.

A dipolar field line satisfies
\beq
r = r_{\rm eq}\sin^2\theta_{\rm m},
\label{eq:dipole_line}
\eeq
where $r_{\rm eq}$ is the radius at which the field line crosses the
magnetic equatorial plane.
For the field line connected to the inner disc edge,
$r_{\rm eq}=R_{\rm m}$.
We therefore describe the central surface of the loaded
magnetospheric flow by
\beq
F(r,\theta_{\rm m})
\equiv
\sin^2\theta_{\rm m}
-
\frac{r}{R_{\rm m}}
=
0.
\label{eq:stream_surface}
\eeq

The local width of the flow changes along the dipolar field line.
Let $\lambda_{\rm m}=\pi/2-\theta_{\rm m}$ be the magnetic latitude.
The angle between the local dipolar field direction and the radial
direction is
\beq
\xi_{\rm f}
=
\arctan\left(\frac{1}{2\tan\lambda_{\rm m}}\right).
\label{eq:field_inclination}
\eeq

Mapping the radial penetration width at the disc inner edge along the
dipolar field line gives the local transverse thickness of the flow,
\beq
H
=
H_{\rm d}
\cos^2\lambda_{\rm m}
\sin\xi_{\rm f}.
\label{eq:Hstream}
\eeq
$H$ is then the characteristic path length travelled by a photon crossing the stream.

The density of the plasma-loaded flow is estimated from the surface density of the material entering the magnetosphere.
We denote this surface density by $\Sigma$, evaluated from the same inner-disc model used to obtain $H_{\rm d}$.
The local mass density in the stream is then approximated as $\rho={\Sigma}/{H}$,
and this density is used to compute the plasma contribution to the dielectric tensor.
This prescription is intended to capture the order of magnitude of the plasma birefringence in the magnetospheric flow, rather than the detailed hydrodynamic structure of an accretion curtain.

During ray tracing, we locate stream crossings by evaluating
$F(r,\theta_{\rm m})$ along each photon trajectory.
A crossing is identified when $F$ changes sign between two successive
integration steps.

At the crossing point we compute the local flow thickness $H$, density
$\rho$, magnetic field, and the incidence angle of the photon on the
stream surface.
Let $\hat{\boldsymbol{k}}$ be the local photon propagation direction
and $\hat{\boldsymbol{n}}_{\rm s}$ the unit normal to the surface
$F=0$.
We define
\beq
\sin\alpha_{\rm s} =
\left|\hat{\boldsymbol{k}}\cdot\hat{\boldsymbol{n}}_{\rm s}\right|,
\label{eq:stream_incidence}
\eeq
where $\alpha_{\rm s}$ is the angle between the photon direction and
the stream surface.

Thus, $\alpha_{\rm s}\to0$ corresponds to a grazing crossing, while
$\alpha_{\rm s}\to\pi/2$ corresponds to a nearly normal crossing.
The effective path length through the plasma layer is then
\beq
l_{\rm s} =
\frac{H}{\sin\alpha_{\rm s}}.
\label{eq:stream_path_length}
\eeq

Scattering opacity, absorption and re-emission inside the stream are
neglected.
This allows us to isolate the birefringent propagation effect of the
plasma: the stream changes the photon polarisation through the phase lag
accumulated between plasma normal modes, not through dilution by
scattered or reprocessed radiation.

\subsection{Vacuum critical density}
\label{app:vacuum_density}

\begin{figure}
\centering
\includegraphics[width=8.5cm]{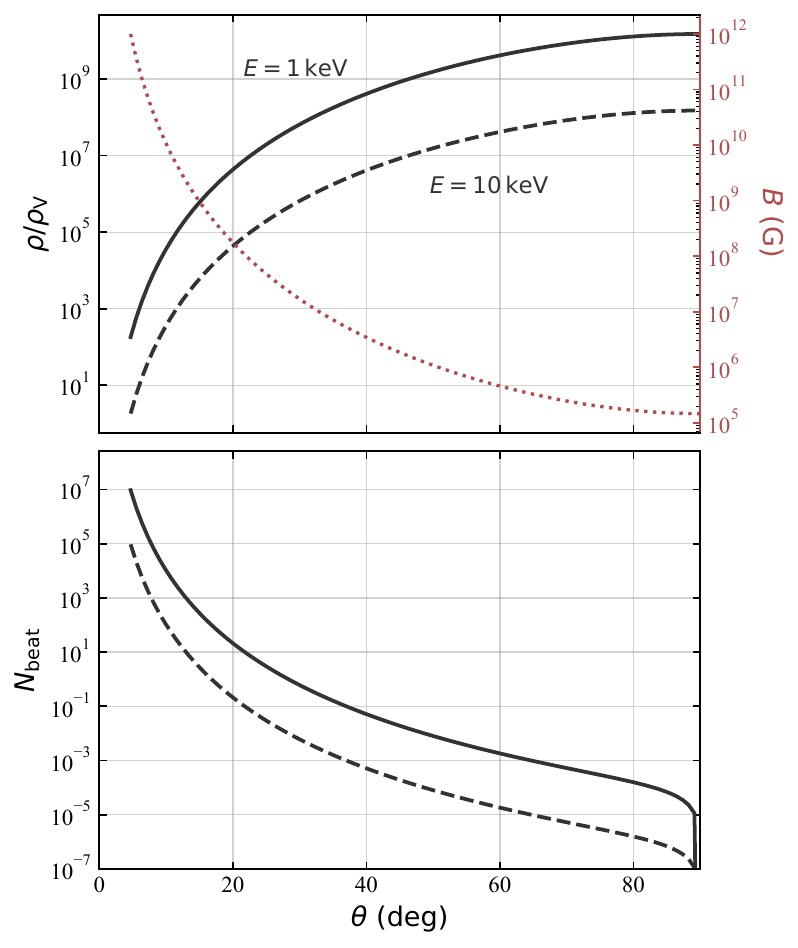}
\caption{
Plasma-dominated propagation in the magnetospheric accretion stream.
Top panel: ratio of the local stream density to the vacuum-resonance
density, $\rho/\rho_{\rm V}$, as a function of magnetic colatitude
$\theta$ for $E=1\,{\rm keV}$ and $E=10\,{\rm keV}$.
The dotted red curve shows the local magnetic-field strength.
Where $\rho/\rho_{\rm V}\gg1$, the normal modes are controlled primarily
by the plasma contribution to the dielectric tensor.
Bottom panel: number of accumulated beat lengths $N_{\rm beat}$ for a
photon crossing the magnetospheric accretion flow.
The associated phase lag between the two plasma normal modes is
$\Delta\Phi=2\pi N_{\rm beat}$, corresponding to a Faraday-like rotation
$\Delta\chi=\pi N_{\rm beat}$ in the circular-mode limit.
The model shown here uses $B_{\rm p}=10^{12}\,{\rm G}$ and
$L=10^{37}\,{\rm erg\,s^{-1}}$.
}
\label{pic:flow_str}
\end{figure}

Assuming a dipolar topology, the local magnetic-field strength is
\beq
\label{eq:Bstream}
B_{\rm loc}
=
\frac{B_{\rm p}}{2}
\left(\frac{R}{r}\right)^3
\left(1+3\cos^2\theta_{\rm m}\right)^{1/2};
\eeq
the corresponding profile is shown by the dotted curve in the upper
panel of Fig.~\ref{pic:flow_str}.

To determine whether propagation inside the flow is plasma- or
vacuum-dominated, we compare the local mass density with the
vacuum-resonance density
\beq
\label{eq:rhoV}
\rho_{\rm V}
=
9.64\times10^{-5}
\left(\frac{B_{\rm loc}}{10^{12}\,{\rm G}}\right)^2
E_{\rm keV}^2
\ {\rm g\,cm^{-3}},
\eeq
where $E_{\rm keV}$ is the photon energy in keV.

When $\rho\gg\rho_{\rm V}$, the plasma contribution dominates the
dielectric tensor and normal modes are predominantly determined by the
plasma response.
In the opposite limit, the polarisation evolution is controlled
primarily by QED vacuum birefringence.
For parameters typical of XRPs, the plasma contribution is dominant
across most of the magnetospheric flow
(upper panel of Fig.~\ref{pic:flow_str}).

\subsection{Plasma normal modes and phase retardation across the stream}
\label{app:plasma_modes}

In the plasma-dominated stream the normal modes are generally different
from the linearly polarised vacuum modes
\citep{1974JETP...38..903G,2003MNRAS.338..233H,2006RPPh...69.2631H}.

We work in the local orthonormal basis where the $z$-axis is aligned
with the photon momentum $\boldsymbol{k}$, the $x$-axis is perpendicular
to the $\boldsymbol{k}$--$\boldsymbol{B}$ plane, and the $y$-axis lies
in that plane.
The wave electric field can then be written as
\begin{equation}
\boldsymbol{E}
=
\left(
E_x\boldsymbol{e}_x
+
E_y\boldsymbol{e}_y
\right)
e^{-i\omega t}.
\end{equation}

For a cold magnetised plasma, neglecting vacuum polarisation, the
ellipticity of the normal modes can be approximated as
\beq
&\xi_\alpha(E,\theta_{\rm kB})
= -i \left(\frac{E_y}{E_x}\right)_\alpha \\
\nonumber
&=
\frac{2\cos\theta_{\rm kB}}
{
\left(\frac{E_{\rm cyc}}{E}\right)\sin^2\theta_{\rm kB}
- (-1)^\alpha
\left[
\left(\frac{E_{\rm cyc}}{E}\right)^2\sin^4\theta_{\rm kB}
+ 4\cos^2\theta_{\rm kB} \right]^{1/2}
},
\label{eq:plasma_xi}
\eeq
where $\theta_{\rm kB}$ is the angle between the photon momentum and
the local magnetic field, $E_{\rm cyc}$ is the local electron cyclotron
energy, and $\alpha=1,2$ labels the two orthogonal plasma modes.

The plasma modes are generally elliptically polarised: they approach
circular polarisation when $|\xi_\alpha|\simeq1$, and become
quasi-linear for propagation nearly perpendicular to the magnetic field.

The two plasma modes propagate with different phase velocities.
The accumulated phase lag is determined by the refractive-index
difference
\beq
\Delta n_{\rm pl}=n_1-n_2.
\eeq

For the flow parameters considered here, it is
$E_{\rm cyc}\ll E$ for typical X-ray photons, because the magnetic field
at the stream crossing points is much smaller than the surface field.
We therefore use the high-frequency limit of the cold
magnetoactive-plasma dielectric tensor in which the refractive-index
splitting is set by the gyrotropic, off-diagonal part of the dielectric
tensor.
Projecting this contribution along the magnetic-field direction gives
the leading-order estimate
\beq
\Delta n_{\rm pl}
\simeq
v\,u_{\rm e}^{1/2}\cos\theta_{\rm kB},
\label{eq:deltan}
\eeq
where
\beq
v=
\left(\frac{E_{\rm p}}{E}\right)^2,
\qquad
u_{\rm e}
=
\left(\frac{E_{\rm cyc}}{E}\right)^2,
\eeq
and
$E_{\rm p}\simeq1.17\,n_{\rm e,21}^{1/2}\,{\rm eV}$ is the electron
plasma energy, with $n_{\rm e}=\rho/m_{\rm p}$ the electron number
density.

Equation~(\ref{eq:deltan}) is the standard high-frequency,
cold-plasma birefringence limit, valid away from the electron cyclotron
resonance
\citep{1974JETP...38..903G,1992herm.book.....M,2003MNRAS.338..233H}.

The beat length for the relative phase accumulation is defined as the
distance over which the relative phase of the two plasma normal modes
changes by $2\pi$,
\beq
l_{\rm beat}
\equiv
\frac{\lambda}{|\Delta n_{\rm pl}|},
\eeq
where
\beq
\lambda = \frac{hc}{E} =
1.24\times10^{-7}
E_{\rm keV}^{-1}\,{\rm cm}
\eeq
is the wavelength.
Thus
\beq
l_{\rm beat} =
\frac{1.24\times10^{-7}}
{E_{\rm keV}|\Delta n_{\rm pl}|} \ {\rm cm}.
\label{eq:lbeat}
\eeq

The effective path length through the stream is
$l_{\rm s}=H/\sin\alpha_{\rm s}$, where $\alpha_{\rm s}$ is defined in
equation~(\ref{eq:stream_incidence}).
The accumulated number of beat lengths is therefore
\beq
N_{\rm beat} =
\frac{l_{\rm s}}{l_{\rm beat}}
= \frac{H}{\sin\alpha_{\rm s}\,l_{\rm beat}}.
\label{eq:Nbeat}
\eeq

The corresponding phase lag between the two plasma modes is then
\beq
\Delta\Phi=2\pi N_{\rm beat}.
\eeq

\subsection{Polarisation update at a stream crossing}
\label{app:stream_update}

The magnetospheric accretion flow is treated as a thin birefringent
layer.
When a ray crosses the stream boundary, we compute the local flow
parameters at the intersection point and update the polarisation state,
accounting for propagation through a layer of thickness $l_{\rm s}$.

The first step is to choose a local polarisation basis at the crossing
point.
Both basis vectors are perpendicular to the photon momentum
$\boldsymbol{k}$.
We take $\boldsymbol{e}_y$ to lie in the plane formed by
$\boldsymbol{k}$ and the local magnetic field $\boldsymbol{B}$, and
$\boldsymbol{e}_x$ to be perpendicular to this plane.
The incident polarisation state is then written in this
$(\boldsymbol{e}_x,\boldsymbol{e}_y)$ basis.

In the plasma-dominated stream, a linearly polarised wave is generally
not one of the propagation eigenmodes.
Instead, it is a superposition of two orthogonal plasma normal modes,
which are in general elliptically polarised and propagate with
different phase velocities.
As the photon crosses the stream, these two mode amplitudes therefore
accumulate a relative phase.
This phase lag changes the polarisation state of the emerging wave; in
the circular-mode limit it appears as a rotation of the linear
polarisation angle.

We decompose the incident state into the two plasma normal modes given
by equation~(\ref{eq:plasma_xi}).
The two mode amplitudes are then multiplied by opposite phase factors,
\beq
a_1
\rightarrow
a_1e^{-i\Delta\Phi/2},
\qquad
a_2
\rightarrow
a_2e^{+i\Delta\Phi/2}.
\label{eq:mode_phase}
\eeq

Finally, the polarisation state is transformed back to the transport
basis used in the ray-tracing calculation.

In the limiting case where the plasma normal modes are nearly circular,
equation~(\ref{eq:mode_phase}) reduces to a Faraday-like rotation of
the linear polarisation angle,
\beq
\Delta\chi
=
\frac{\Delta\Phi}{2}
=
\pi N_{\rm beat}.
\label{eq:faraday_limit}
\eeq

In general, however, the plasma modes are elliptically polarised.
The stream then does not merely rotate the linear polarisation angle:
it can also convert linear into circular polarisation and vice versa.
For this reason the calculation is formulated in terms of the Stokes,
or equivalently Jones, vector rather than as a scalar rotation of the
polarisation angle.

\begin{figure*}
\centering
\includegraphics[width=16.cm]{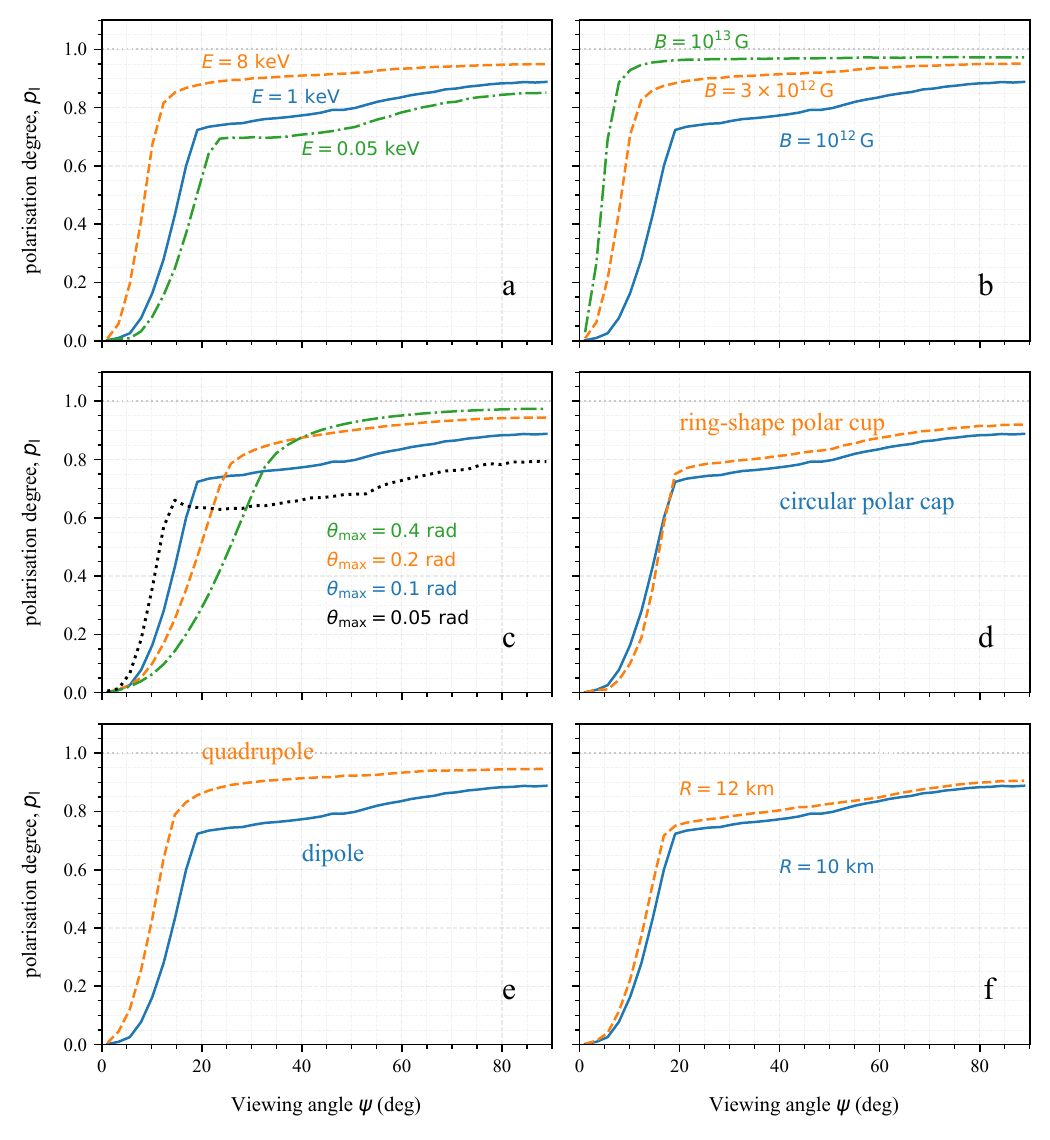}
\caption{
Hotspot-integrated linear polarisation degree $p_{\rm l}$ versus
viewing angle $\psi$ for the vacuum/QED propagation channel.
Seed emission is fully linearly polarised ($p_{{\rm l},i}=1$) from two
antipodal caps.
(a) $E=0.05$, $1$ and $8\,\mathrm{keV}$;
(b) $B=10^{12}$, $3\times10^{12}$ and $10^{13}\,\mathrm{G}$;
(c) $\theta_{\max}=0.05$, $0.1$, $0.2$ and $0.4\,\mathrm{rad}$;
(d) filled circular cap versus ring-shaped polar cap;
(e) dipole versus quadrupole field;
(f) $R=10$ and $12\,\mathrm{km}$ at fixed $M=1.4\,M_\odot$.
The calculations include QED vacuum birefringence and the associated
local non-adiabatic pockets, but do not include propagation through the
magnetospheric accretion flow.
Fiducial parameters (when not varied):
$B=3\times10^{12}\,\mathrm{G}$,
$E=1\,\mathrm{keV}$,
$\theta_{\max}=0.1\,\mathrm{rad}$,
$M=1.4\,M_\odot$,
$R=10\,\mathrm{km}$.
}
\label{pic:p_l}
\end{figure*}

\subsection{Coupling to the QED adiabatic zone}
\label{app:stream_qed_coupling}

The observable effect of a flow crossing depends on where it occurs
relative to the QED adiabatic region.
If the photon crosses the stream after its polarisation has frozen out,
the stream mainly acts as a birefringent retarder; in the circular-mode
limit this appears as a rotation of the observed polarisation position
angle by $\Delta\chi$, given by equation~(\ref{eq:faraday_limit}).

However, when the photon crosses the stream while still inside the QED
adiabatic region, the plasma layer rotates the polarisation state
relative to the local QED eigenmode basis.
After leaving the stream, adiabatic propagation in the QED-dominated
magnetosphere resumes and the photon is re-directed into the local
ordinary and extraordinary modes.
This projection converts the stream-induced phase offset into partial
mode mixing.

Thus, for crossings inside the adiabatic zone, the accretion stream
provides an additional single-ray depolarisation channel, analogous to
the depolarisation produced at the boundaries of non-adiabatic pockets.

In the simplified linear-mode limit, if a photon entering the stream is
initially an incoherent mixture with fractions $f_{\rm O}$ and
$f_{\rm X}=1-f_{\rm O}$, and the stream rotates the linear polarisation
angle by $\Delta\chi$, the fractions after projection onto the local
O/X basis become
\beq
f_{\rm O}'
=
f_{\rm O}\cos^2\Delta\chi
+
f_{\rm X}\sin^2\Delta\chi ,
\label{eq:fo_update}
\eeq
\beq
f_{\rm X}'
= f_{\rm O}\sin^2\Delta\chi +
f_{\rm X}\cos^2\Delta\chi .
\label{eq:fx_update}
\eeq

The corresponding single-ray linear polarisation degree is
\beq
p_{\rm l}
=
\frac{|f_{\rm O}'-f_{\rm X}'|}
{f_{\rm O}'+f_{\rm X}'} .
\label{eq:single_ray_stream_pd}
\eeq

\section{Additional dependence on the model parameters}
\label{app:parameter_dependence}

The dependence of the QED propagation effect on the principal physical and geometrical parameters is illustrated in Fig.~\ref{pic:p_l}.
The figure shows the hotspot-integrated linear polarisation degree for different photon energies, magnetic-field strengths, emitting-region sizes and shapes, magnetic-field configurations, and NS compactness.

\subsection{Characteristic emitting-region size for the sign change
of the QED effect}
\label{app:theta_crit}

The net effect of QED vacuum birefringence on the hotspot-integrated
linear polarisation degree depends on the angular extent of the
emitting region.

For sufficiently small emitting regions, local non-adiabatic
propagation pockets imprint ray-dependent polarisation-angle offsets
that lead to an additional depolarisation upon summation.
For extended emitting regions, surface averaging already suppresses
the polarisation in the absence of QED, and vacuum birefringence instead
tends to partially re-align the polarisation vectors before freeze-out.

As a result, the sign of the QED contribution to the observable linear
polarisation changes at a characteristic polar-cap half-opening angle,
$\theta_{\rm crit}$, defined such that the net polarisation degree
computed with and without QED vacuum birefringence is approximately
the same.

For a dipolar magnetic field, we find that $\theta_{\rm crit}$ can be
approximated by the empirical expression
\beq\label{eq:theta_crit_app}
\theta_{\rm crit}
\approx
0.04 + 0.063\,E_{\rm keV}^{-1.7}
+ 0.17\,E_{\rm keV}^{-0.45}B_{12}^{-0.4}
\,{\rm rad},
\eeq
where $E_{\rm keV}$ is the photon energy in keV and
$B_{12}=B/(10^{12}\,\mathrm{G})$ is the surface magnetic-field strength.
This approximation is valid over the parameter range
\beq
10^{10}\,\mathrm{G}
\lesssim
B
\lesssim
10^{14}\,\mathrm{G},
\qquad
1\,\mathrm{keV}
\lesssim
E
\lesssim
32\,\mathrm{keV},
\eeq
which encompasses the conditions relevant for XRPs observed with IXPE.

Equation~(\ref{eq:theta_crit_app}) provides a compact diagnostic for
assessing whether magnetospheric QED propagation is expected to
decrease or increase the observable linear polarisation degree for a
given emitting-region size, photon energy, and magnetic-field strength.
In the XRP regime, the resulting values of $\theta_{\rm crit}$
correspond to genuinely compact emission regions, typically of the
order of a few degrees.

\section{Scattering and absorption in the magnetospheric accretion flow}
\label{app:scattering}

\begin{figure}    
\label{fig:unscattered_fraction}
    \centering
    \includegraphics[width=\columnwidth]{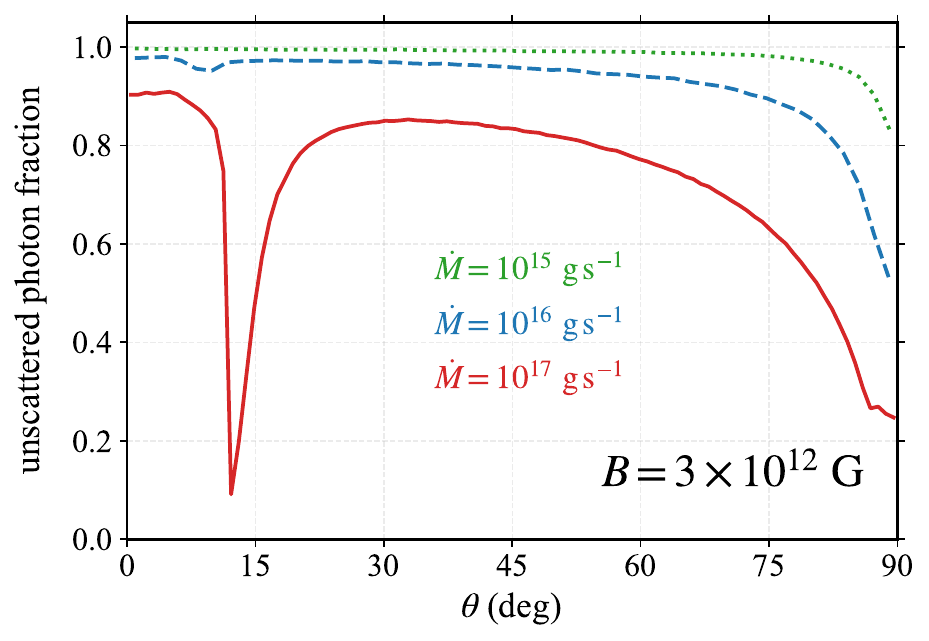}
    \caption{
    Fraction of escaping photons that experience no scattering,
    $f_0$, as a function of the magnetic zenith angle $\theta_{\rm m}$.
    The calculations assume a dipolar magnetic field with
    $B=3\times10^{12}\ {\rm G}$ and mass accretion rates
    $\dot M=10^{15}$, $10^{16}$, and $10^{17}\ {\rm g\,s^{-1}}$.
    Each Monte Carlo simulation contains $10^{6}$ photons
    emitted from the polar footprint. At
    $\dot M\lesssim10^{16}\ {\rm g\,s^{-1}}$ the escaping radiation
    remains dominated by unscattered photons over most viewing
    directions, whereas scattering becomes important over a broad
    angular range at $\dot M=10^{17}\ {\rm g\,s^{-1}}$.
    }
\end{figure}

The polarisation calculations presented in this work treat the magnetospheric accretion flow as a birefringent medium and neglect photon scattering and true absorption. 
Here we test the validity of this approximation for the sub-critical accretion regime considered in this work.

True absorption is negligible under these conditions. 
As an order-of-magnitude estimate, the non-magnetic free--free absorption coefficient in a fully ionised hydrogen plasma is
\beq
    \alpha_{\rm ff}\simeq
    3.7\times10^{8}\,
    T^{-1/2} n_{\rm e}^{2} \nu^{-3}
    \left(1-e^{-h\nu/kT}\right) g_{\rm ff}
    \quad {\rm cm^{-1}}.
\eeq
For the densities characteristic of the magnetospheric flow and photon energies in the keV range, the corresponding absorption optical depth is much smaller than unity. 
Magnetic corrections modify the detailed mode- and angle-dependent opacity but do not change this conclusion. 
We therefore neglect true absorption and consider explicitly only photon scattering.

We test the importance of scattering with Monte Carlo simulations of photon propagation through the magnetospheric accretion flow. 
The flow geometry and density distribution are calculated following the prescription of \citet{2026MNRAS.tmp.1203M}. 
For each model, $10^{6}$ photons are emitted from the polar footprint and propagated through the magnetosphere, accounting for GR effects. 
The scattering probability is calculated from the local plasma density. 
Following each scattering, the photon is propagated until it encounters the flow or the stellar surface again, or escapes from the magnetosphere.

As a direct measure of the importance of scattering, we calculate
\beq
    f_0(\theta_{\rm m})=
    \frac{N_{\rm esc}(N_{\rm sc}=0)}
         {N_{\rm esc}},
\eeq
where $f_0$ is the fraction of photons escaping at magnetic zenith angle $\theta_{\rm m}$ without experiencing any scattering. 
The results for $B=3\times10^{12}\ {\rm G}$ and three mass accretion rates are shown in Fig.~\ref{fig:unscattered_fraction}.

At $\dot M=10^{15}$ and $10^{16}\ {\rm g\,s^{-1}}$, the escaping radiation is dominated by unscattered photons over most viewing directions. 
In particular, $f_0$ remains close to unity towards the magnetic axis, i.e. in the directions most relevant to the propagation effects discussed in this work. The decrease of $f_0$ towards the magnetic equatorial plane is enhanced by photons redirected by scattering and reflection from the stellar surface.

The situation changes at $\dot M=10^{17}\ {\rm g\,s^{-1}}$, where scattering becomes important over a much broader range of viewing angles and the direct component can be substantially reduced. 
At such accretion rates the source is already approaching the transition from sub-critical hotspot accretion towards the formation of a radiation-dominated accretion column, for which both the assumed emission geometry and the neglect of radiative transfer through the flow become questionable.

Thus, we conclude that photon scattering does not qualitatively affect the propagation picture developed in this work at sufficiently low sub-critical accretion rates, $\dot M\lesssim10^{16}\ {\rm g\,s^{-1}}$. 
At accretion rates approaching $\sim10^{17}\ {\rm g\,s^{-1}}$, however, scattering can no longer be neglected and should be treated together with polarisation transport in a self-consistent radiative-transfer calculation.

\bsp 
\label{lastpage}
\end{document}